\documentclass[conference]{IEEEtran} 
\IEEEoverridecommandlockouts
\usepackage[binary-units,per-mode=symbol,per-symbol=p]{siunitx}
\usepackage{cite}
\usepackage{adjustbox}
\usepackage[ruled,vlined]{algorithm2e}
\usepackage{amsmath,amssymb,amsfonts}
\usepackage{algorithmic}
\usepackage{graphicx}
\usepackage{textcomp}
\usepackage[colorlinks=false, pdfborder={0 0 0}, breaklinks]{hyperref}
\usepackage[inline]{enumitem}

\usepackage{multirow}
\usepackage{multicol}
\usepackage{subcaption}
\def\BibTeX{{\rm B\kern-.05em{\sc i\kern-.025em b}\kern-.08em
    T\kern-.1667em\lower.7ex\hbox{E}\kern-.125emX}}

\usepackage{amsmath}
\DeclareMathOperator*{\argmax}{arg\,max}

\usepackage{mathtools}

\usepackage{tikz}
\usetikzlibrary{shapes.geometric, arrows, positioning}

\usepackage{ctable}
\usepackage{balance}

\newcommand{\ie}{\emph{i.e.}, }
\newcommand{\eg}{\emph{e.g.}, }

\newcommand{\cf}{\emph{cf.}\xspace}

\newcommand{\arcs}{\texttt{ARCS}\xspace}

\newcommand{\BDRP}{BDR\textsubscript{P}}

\newcommand{\BDRC}{BDR\textsubscript{C}}

\tikzstyle{frame} = [rectangle, rounded corners, minimum width=3cm, minimum height=1cm,text centered, draw=black, fill=cyan!20]
\tikzstyle{cnn} = [rectangle, rounded corners, minimum width=4.5cm, minimum height=1cm,text centered, draw=black, fill=blue!20]
\tikzstyle{dense} = [rectangle, rounded corners, minimum width=4.5cm, minimum height=1cm,text centered, draw=black, fill=orange!20]
\tikzstyle{meta} = [rectangle, rounded corners, minimum width=3cm, minimum height=1cm,text centered, draw=black, fill=green!20]
\tikzstyle{output} = [rectangle, rounded corners, minimum width=3.5cm, minimum height=1cm,text centered, draw=black, fill=lime!20]
\tikzstyle{arrow} = [thick,->,>=stealth]

\begin{document}

\title{Adaptive Resolution and Chroma Subsampling for Energy-Efficient Video Coding}

\author{\IEEEauthorblockN{Amritha Premkumar\IEEEauthorrefmark{1}, Christian Herglotz\IEEEauthorrefmark{1}
}
\IEEEauthorblockA{\IEEEauthorrefmark{1}Chair of Computer Engineering, Brandenburg University of Technology Cottbus-Senftenberg, Germany}
}

\maketitle

\begin{abstract}
Conventional video encoders typically employ a fixed chroma subsampling format, such as YUV420, which may not optimally reflect variations in chroma detail across different types of content. This can lead to suboptimal chroma quality and inefficiencies in bitrate allocation. We propose an Adaptive Resolution–Chroma Subsampling (\arcs) framework that jointly optimizes spatial resolution and chroma subsampling to balance perceptual quality and decoding efficiency. \arcs selects an optimal (resolution, chroma format) pair for each bitrate by maximizing a composite quality–complexity objective, while enforcing monotonicity constraints to ensure smooth transitions between representations. Experimental results using x265 show that, compared to a fixed-format encoding (YUV444), on average, \arcs achieves a \SI{13.48}{\percent} bitrate savings and a \SI{62.18}{\percent} reduction in decoding time, which we use as a proxy for the decoding energy, to yield the same colorVideoVDP score. The proposed framework introduces chroma adaptivity as a new control dimension for energy-efficient video streaming.
\end{abstract}

\begin{IEEEkeywords}
Chroma subsampling, Energy-efficient video coding, Adaptive streaming, HEVC, Rate–distortion, Perceptual quality
\end{IEEEkeywords}

\section{Introduction}
The global demand for video streaming continues to grow, accounting for the majority of consumer Internet traffic~\cite{CiscoForecast}. As a result, reducing the energy footprint of video delivery has become an urgent research challenge, driven by both environmental concerns and the limited energy budgets of mobile and embedded devices~\cite{chachou_energy_2023, Herglotz_hevc_deceenrgy_time, Herglotz_hevc_hardware_decoder_energy, Herglotz_energy_reduce_hdr_opportunity, Mercat_10bit_hevc_energy_hdr}. Recent works have explored multiple avenues to improve energy efficiency in HTTP Adaptive Streaming (HAS), including content-aware spatial resolution adaptation~\cite{gnostic, netflix_paper, jtps_ref, ladre}, variable framerate encoding~\cite{cvfr_ref, rajendran_energy-quality-aware_2024}, and efficient per-title bitrate ladder construction~\cite{emes}.

While these approaches have shown substantial benefits, they share a common limitation: \emph{the chroma representation of video content is almost always treated as fixed}. In practice, most adaptive streaming pipelines use YUV420 chroma subsampling as the default format due to its broad hardware support and lower bitrate requirements, even though modern codecs support more expressive chroma subsampling (\eg HEVC “Main” profiles default to YUV420). This convention persists despite extensive research in chroma processing for High Dynamic Range (HDR), Wide Color Gamut (WCG), and Standard Dynamic Range (SDR) content—such as weighted chroma downsampling and luma-referenced upsampling~\cite{fu_weighted_chroma_downsample}, subjective and objective evaluations of chroma subsampling for HDR~\cite{Boitard_chroma_hdr, Thoma_chroma_hdr}, and early very-low-bitrate chroma coding strategies~\cite{Karam_chroma_coding}.

\begin{figure}[t]
\centering
\includegraphics[width=0.45\linewidth]{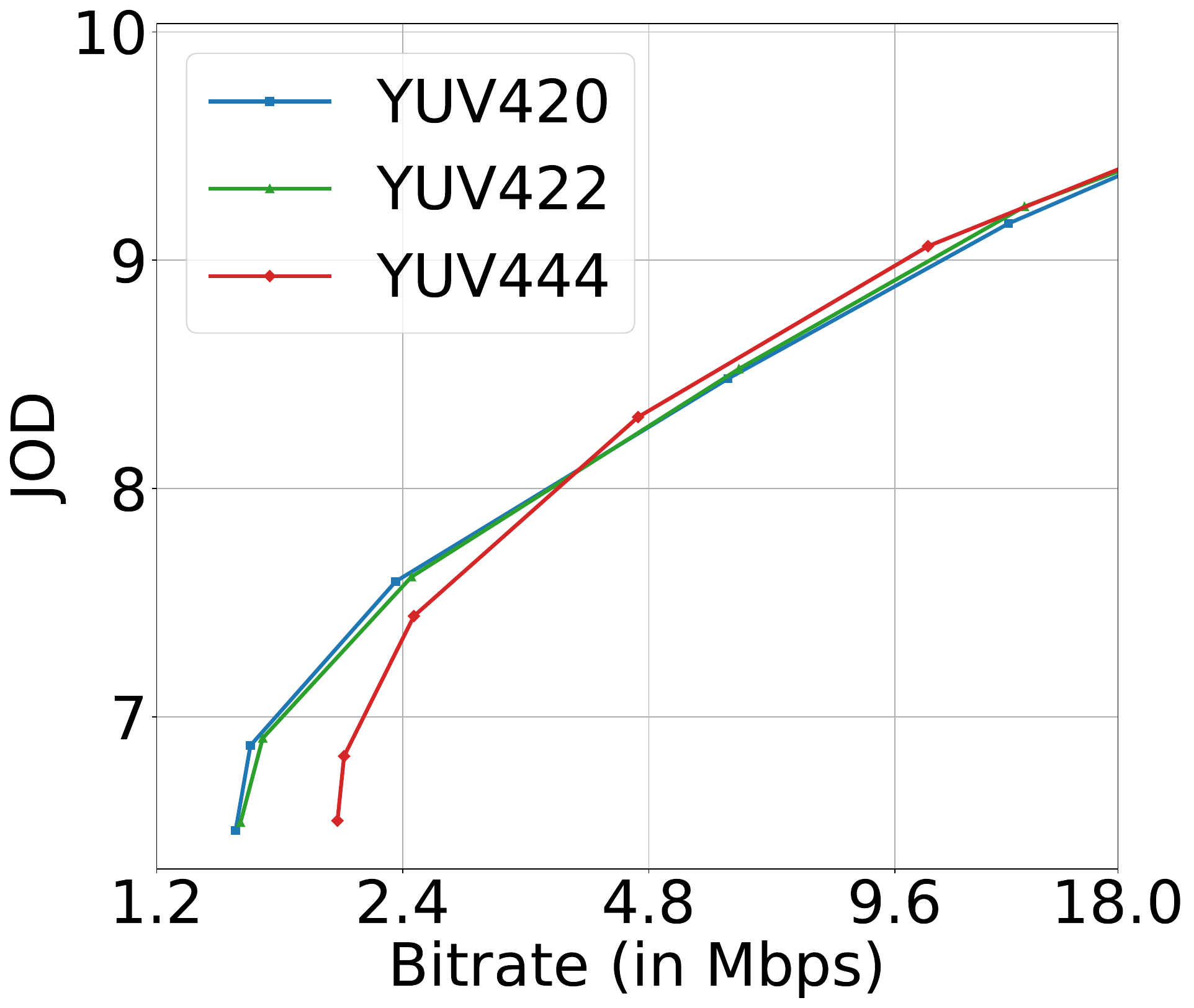}
\includegraphics[width=0.45\linewidth]{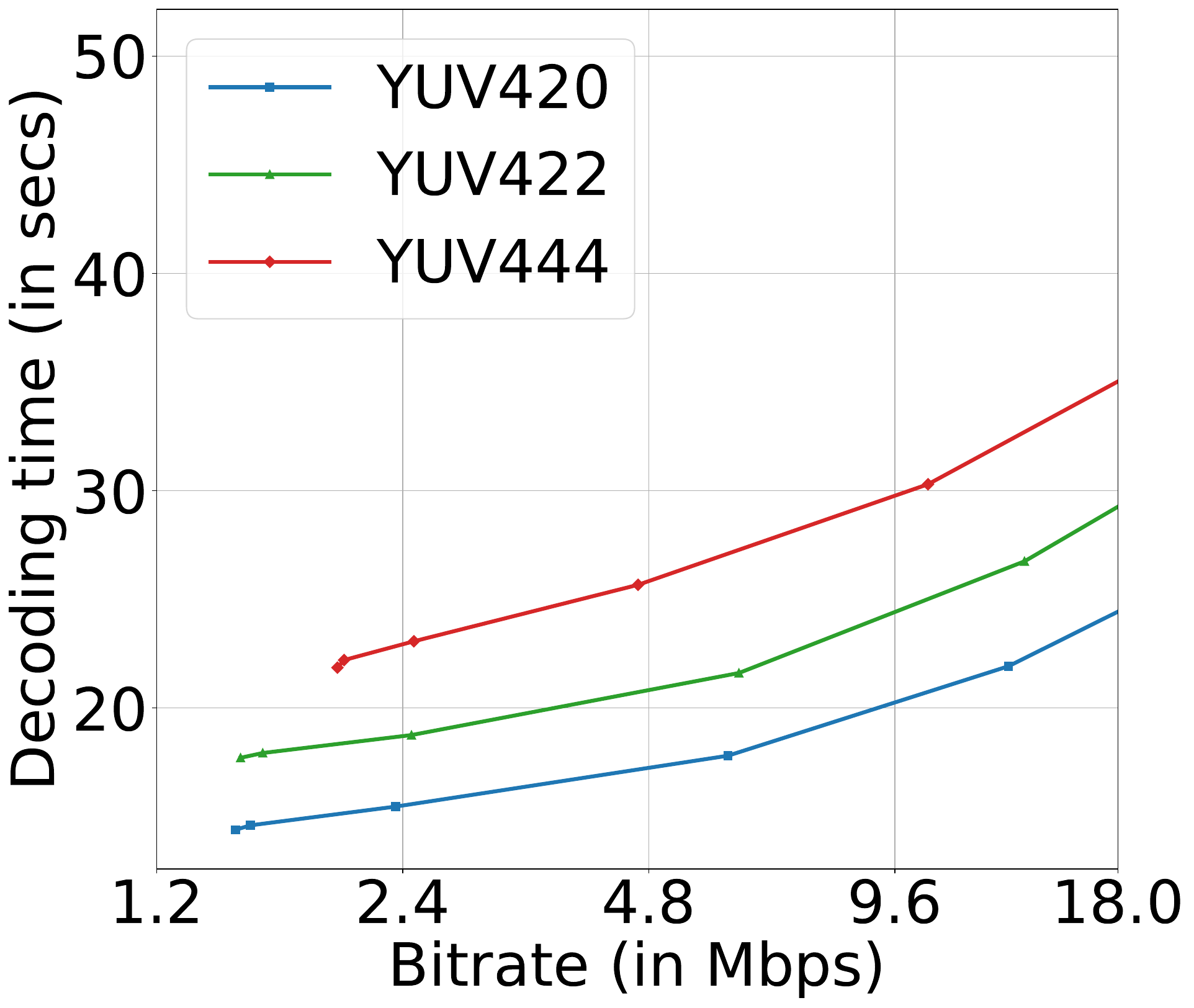}

\caption{Rate–Quality and Rate–Time curves for the \emph{Campfire Party} sequence~\cite{sjtu_video_ref} encoded with x265 at ultra high definition UHD using YUV444, YUV422, and YUV420 formats. ``Time'' denotes decoding time, and ``Quality'' denotes colorVideoVDP in Just Objectionable Difference (JOD) units~\cite{ColorVideoVDP_ref}.}

\vspace{-1em}
\label{fig:rqt_paramspace}
\end{figure}

However, not all content exhibits the same chroma complexity. The human visual system (HVS) exhibits distinct chromatic processing in the cortex and varying color sensitivity under specific conditions—such as low luminance levels (where chroma sensitivity decreases) or high-saturation regions (where it increases)~\cite{color_ref1, color_ref2}. Recent HVS-inspired perceptual color compression methods exploit these effects to improve perceptual efficiency~\cite{color_ref3}.
Fig.~\ref{fig:rqt_paramspace} illustrates the aggregate rate–quality–complexity behavior of a representative sequence of the SJTU 4K dataset~\cite{sjtu_video_ref} across YUV444, YUV422, and YUV420 formats. It is observed that YUV444 videos require up to twice the decoding time as YUV420, while YUV420 has higher quality than YUV444 at bitrates up to 3.6\,Mbps. This observation suggests that chroma processing accounts for a significant share of overall decoding cost, yet much of this computation is redundant in low-color or desaturated scenes. Consequently, \emph{chroma adaptivity remains an underexplored opportunity} for reducing energy consumption without perceptual compromise.

Recent studies, such as~\cite{Amirpour_bitdepth_chroma_study}, have examined the impact of bit depth and color subsampling on coding efficiency and energy use across SDR and HDR content. While these analyses provide valuable insights into the trade-offs between 8-bit and 10-bit formats and fixed chroma subsampling formats (YUV444, YUV422, YUV420), they rely on brute-force per-title evaluations of all combinations. This static formulation does not incorporate adaptivity and remains isolated from encoder control logic or adaptive streaming frameworks, what we refer to as the \emph{codec-side decision pipeline}, where encoding parameters are selected during compression based on content statistics and system constraints. By integrating chroma adaptivity into this pipeline, our method can directly influence encoding decisions and be deployed in practical streaming systems.

To address this research gap, we formulate our central hypothesis: \emph{adaptive chroma subsampling can reduce bitrate and energy consumption without perceptual degradation}. We propose a complexity-driven framework that dynamically selects between spatial resolution and chroma subsampling formats to maximize a composite quality–complexity objective at each target bitrate. By avoiding redundant chroma processing in low-complexity regions, our method achieves significant savings in decoding energy compared to the benchmark approaches. Moreover, it is codec-agnostic and compatible with existing video coding standards such as HEVC~\cite{HEVC}, VVC~\cite{VVC}, and AV1~\cite{han_technical_2021}. Modern codecs already include chroma-related tools, such as luma mapping with chroma scaling (LMCS)~\cite{lu_luma_mapping_chroma_scale_vvc}, adaptive color transforms~\cite{jhu_adaptive_color_transform_vvc}, cross-component and deep-learning-based chroma prediction~\cite{Astola_crosscomponentmodel_chromapred, zhu_dlbased_chromapred_vvc}, and fast chroma intra-mode decision techniques~\cite{Wang_vvc_chroma_mode}, which provide a foundation for integrating adaptive chroma decisions.

\section{Adaptive Resolution Chroma Subsampling (\arcs) Framework}
\label{sec:imp}
\begin{figure}[t]
\centering
\includegraphics[width=1\linewidth]{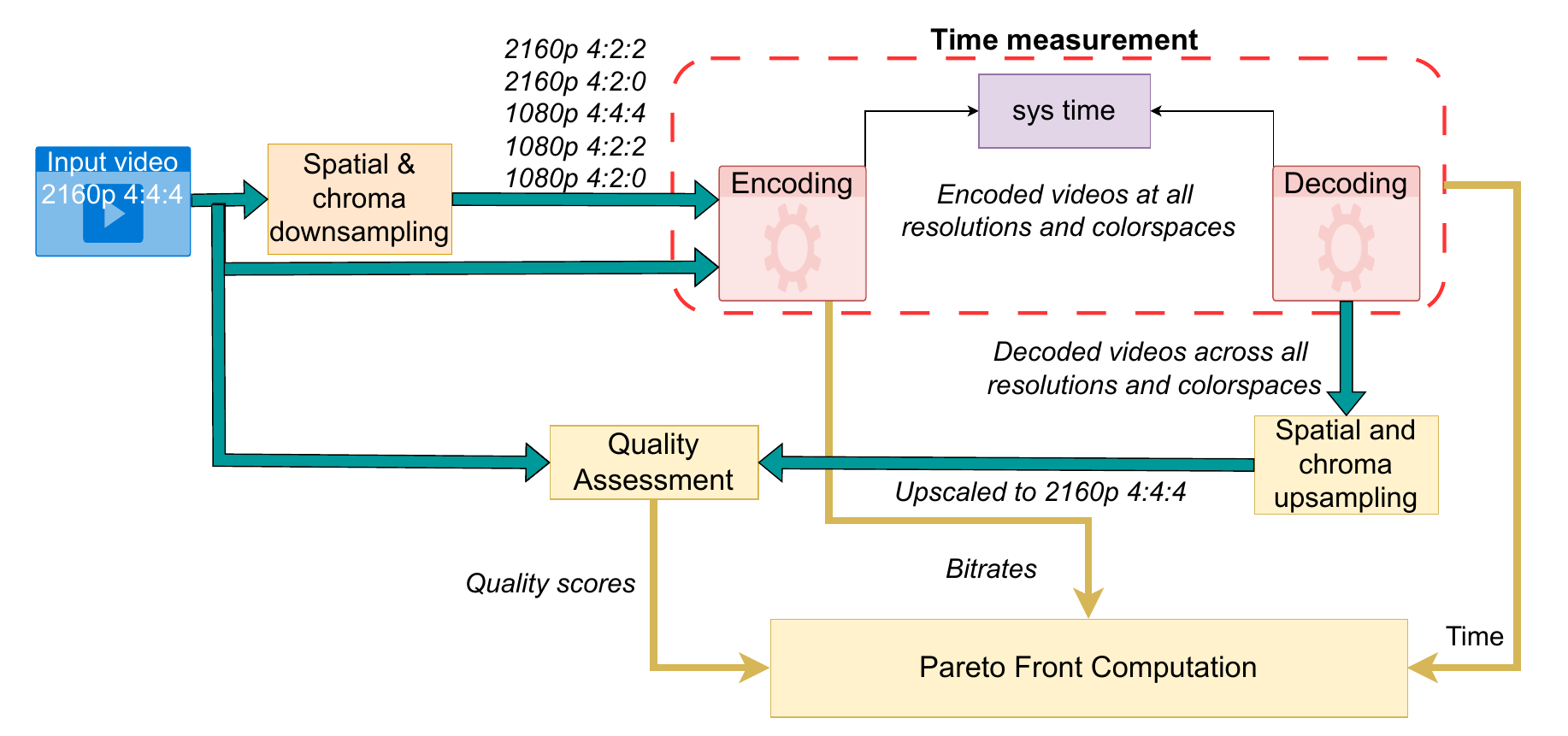}
\caption{Overview of the \arcs methodology.}
\vspace{-0.9em}
\label{fig:sys_arch}
\end{figure}

To enable energy-efficient video delivery while preserving perceptual quality, we propose an adaptive chroma subsampling framework that jointly optimizes \emph{spatial resolution} and \emph{chroma subsampling format}. Instead of treating chroma subsampling as a fixed parameter, the proposed methodology dynamically determines the most suitable chroma representation for each bitrate–resolution pair based on content characteristics and decoding complexity. The overall workflow is illustrated in Fig.~\ref{fig:sys_arch}.

\subsection{Generation of subsampled videos}
The process begins with the original high-quality source video, typically in YUV444 format. The input is spatially and chromatically subsampled to generate multiple candidate representations with different resolutions (\eg 360p–2160p) and chroma subsampling formats (\eg YUV444, YUV422, YUV420). Each chroma format represents a distinct trade-off between color fidelity and sample density. High-quality resampling filters are used to avoid aliasing and preserve structural consistency across all versions.

\subsection{Encoding and Decoding Evaluation}
Each $(r, c)$ configuration, defined by its resolution $r$ and chroma subsampling format $c$, is encoded using a state-of-the-art video encoder and decoded with the corresponding video decoder. For every configuration, we record three key attributes:
\begin{itemize}[leftmargin=*]
    \item \emph{Bitrate:} The output bitrate obtained from the encoder at a given quantization parameter or target bitrate.
    \item \emph{Perceptual Quality:} Objective visual quality metrics such as Video Multi-Method Assessment Fusion (VMAF)~\cite{blog_vmaf_2018}, Peak Signal to Noise Ratio (PSNR)~\cite{yuv_ref}, or ColorVideoVDP~\cite{ColorVideoVDP_ref} are computed after decoding and upscaling, using a common reference domain (\eg 2160p~YUV444) for fair comparison. In this paper, we use ColorVideoVDP as the metric, as it is the first color-aware metric that accounts for both spatial and temporal aspects of vision~\cite{ColorVideoVDP_ref}.
    \item \emph{Complexity:} The decoding time per frame, representing the computational load and serving as a proxy for decoding energy consumption~\cite{Herglotz_hevc_deceenrgy_time}.
\end{itemize}
These measurements form a comprehensive dataset describing the rate–quality–complexity relationship across resolution-chroma pairs.

\subsection{Quality–Complexity Modeling}
To quantify the trade-off between perceptual quality and computational cost, we introduce a composite scalar metric:
\begin{equation}
    J = Q - \alpha \log(\tau_{\text{D}}),
    \label{eq:Jmetric_framework}
\end{equation}
where \( Q \) denotes the perceptual quality score, \( \tau_{\text{D}} \) is the average decoding time per frame, and \( \alpha \) is a tunable coefficient that balances quality and energy efficiency~\cite{katsenou2026multiobjectiveparetofrontoptimizationefficient}. The logarithmic term ensures diminishing penalties for moderate increases in decoding time, allowing refined trade-offs.  

Before optimization, both \( Q \) and \( \log(\tau_{\text{D}}) \) are normalized to the range \([0, 1]\):
\begin{equation}
    J' = \frac{Q - Q_{\min}}{Q_{\max} - Q_{\min}} - 
    \alpha \frac{\log(\tau_{\text{D}}) - \log(d_{\min})}{\log(d_{\max}) - \log(d_{\min})}.
    \label{eq:normalizedJ_framework}
\end{equation}
This normalization avoids bias toward particular scales of quality or complexity and improves generality across content types and devices.

\subsection{Adaptive Decision Optimization}
For each bitrate $b$ in the set of target bitrates ($\mathcal{B}$), the optimal combination of spatial resolution \( r \) and chroma subsampling format \( c \) is determined by maximizing the composite objective:
\begin{equation}
    (r^*, c^*)_{\alpha} = \argmax_{r \in \mathcal{R}, \, c \in \mathcal{C}} J'(r, c, b),
    \label{eq:joint_opt_framework}
\end{equation}
where \( \mathcal{R} \) and \( \mathcal{C} \) denote the sets of available resolutions and chroma subsampling formats, respectively. The optimization selects the representation that provides the best perceptual quality for a given bitrate while minimizing decoding effort, based on the composite scalar metric. The resulting sequence of optimal $(r^*, c^*)$ pairs across all bitrate targets forms a \emph{joint resolution–chroma subsampling optimized bitrate ladder}.

To ensure practical and perceptually consistent adaptation, two additional constraints are introduced:
\begin{itemize}[leftmargin=*]
    \item \emph{Resolution monotonicity:} the selected spatial resolution must be non-decreasing with increasing target bitrate, \ie \( r_{i+1} \geq r_i \). This prevents undesirable downscaling at higher bitrates and aligns with conventional bitrate ladder behavior in adaptive streaming.
    \item \emph{Chroma monotonicity with refresh:} within a fixed resolution segment, chroma fidelity is also constrained to be non-decreasing across higher bitrates, \ie \( c_{i+1} \geq c_i \). However, when the resolution increases (\( r_{i+1} > r_i \)), the chroma level is allowed to \emph{refresh}—that is, it can revert to a lower subsampling format and then continue increasing monotonically for subsequent bitrates.
\end{itemize}

Together, these monotonicity rules yield a smooth and interpretable bitrate ladder that avoids abrupt quality transitions, ensuring both coding stability and perceptual coherence across bitrate levels.

\section{Evaluation Setup}

\subsection{Dataset}
To evaluate the \arcs framework, we use 15 UHD videos from the SJTU dataset~\cite{sjtu_video_ref}, as we require source videos with high chroma fidelity, specifically in the YUV444 format. Such material allows us to subsample to other chroma subsampling formats (YUV422 and YUV420) without introducing preprocessing artifacts that might bias the results.

\subsection{Metrics}
\emph{Quality:} We compute weighted YUV-PSNR~\cite{yuv_ref}, and ColorVideoVDP~\cite{ColorVideoVDP_ref}. ColorVideoVDP reports the video quality in the Just-Objectionable-Difference (JOD) units. All objective scores are computed against the original YUV444 reference after mapping decoded outputs to a \emph{common} reference domain (2160p YUV444) to ensure comparability across resolutions and chroma subsampling formats.

\emph{Decoding complexity:} We measure average decoding time per frame (wall-clock), and use it as a proxy for decoding energy~\cite{Herglotz_hevc_deceenrgy_time}.

For aggregate comparisons, we report Bjøntegaard delta rates~\cite{DCC_BJDelta}, \BDRP,  \BDRC~(BD-rate vs. PSNR/CVVDP), and Bjøntegaard delta decoding-times (BDDT)~\cite{Herglotz_hevc_hardware_decoder_energy} computed over the same operating points using piecewise cubic interpolation on log-rate.

\subsection{Protocol and Implementation Details}
All experiments are run on 2x AMD EPYC
ROME 7352 Processor (48 cores), 512 GB RAM, operating on Alma Linux 9.4. As shown in Table~\ref{tab:exp_par}, for each sequence and target bitrate $b \in \mathcal{B}$, we generate all $(r,c)$ combinations with $r\!\in\!\{1080\text{p},\,2160\text{p}\}$ and $c\!\in\!\{\text{YUV420},\,\text{YUV422},\,\text{YUV444}\}$. A strict tolerance window of $\pm\!10\%$ around each target bitrate in $\mathcal{B}$ is enforced during selection. For each target bitrate, \arcs selects $(r^*,c^*)$ maximizing $J'=Q'-\alpha\,d'$, where $Q'$ and $d'$ are per-title normalized quality and $\log(\tau_\text{D})$ (\cf Eq.~\ref{eq:normalizedJ_framework}). We sweep $\alpha \in [0,1]$ to expose the quality–energy trade-off. We encode with \emph{x265 v4.1} (\texttt{slower} preset) and decode with the HM HEVC reference decoder (single thread). Decoding time excludes I/O, is averaged per frame, and each operating point is measured three times; we report the mean.   

\begin{table}[t]
\caption{Experimental parameters used in this paper.}
\centering
\resizebox{0.895\linewidth}{!}{
\begin{tabular}{l|c|c|c|c|c|c}
\specialrule{.12em}{.05em}{.05em}
\specialrule{.12em}{.05em}{.05em}
\multicolumn{2}{c|}{\emph{Parameter}} & \multicolumn{5}{c}{\emph{Values}}\\
\specialrule{.12em}{.05em}{.05em}
\specialrule{.12em}{.05em}{.05em}
Set of resolutions & $\mathcal{R}$ & \multicolumn{5}{c}{\{ 1080, 2160 \} } \\
\hline
Set of bitrates & $\mathcal{B}$ & 600 & 900 & 1600 & 2400 & 3400  \\
& & 4500 & 5800 & 8100 & 11600 & 16800\\
\hline
Set of colorspaces & $\mathcal{C}$ & \multicolumn{5}{c}{\{YUV420, YUV422, YUV444 \} } \\
\hline
\multicolumn{2}{c|}{\emph{Encoder}} & \multicolumn{5}{c}{\emph{x265 v4.1 [slower]}}\\
\hline
\multicolumn{2}{c|}{\emph{Decoder}} & \multicolumn{5}{c}{\emph{HM}}\\
\specialrule{.12em}{.05em}{.05em}
\specialrule{.12em}{.05em}{.05em}
\end{tabular}
}
\vspace{-0.93em}
\label{tab:exp_par}
\end{table}

\subsection{Benchmarks}
\arcs is compared against the following state-of-the-art methods:
\begin{enumerate}
    \item \emph{FixedLadder}~\cite{HLS2}: We use the HLS bitrate ladder specified by Apple as the fixed set of bitrate-resolution pairs. 
    \item \emph{Default}: We only encode at the native, 2160p, resolution for a given set of bitrates. This method demonstrates a limitation in covering the whole range of the bitrate ladder, especially the low bitrates. 
    \item \emph{DynResJOD}: Ablation study of the colorspace optimization of \arcs. In this method, we optimize only spatial resolution.
\end{enumerate}

\section{Experimental Results}

\subsection{Rate–Quality–Complexity Behavior}
\begin{figure}[t]
\centering
\begin{subfigure}{0.24\textwidth}
\centering
\includegraphics[width=1\textwidth]{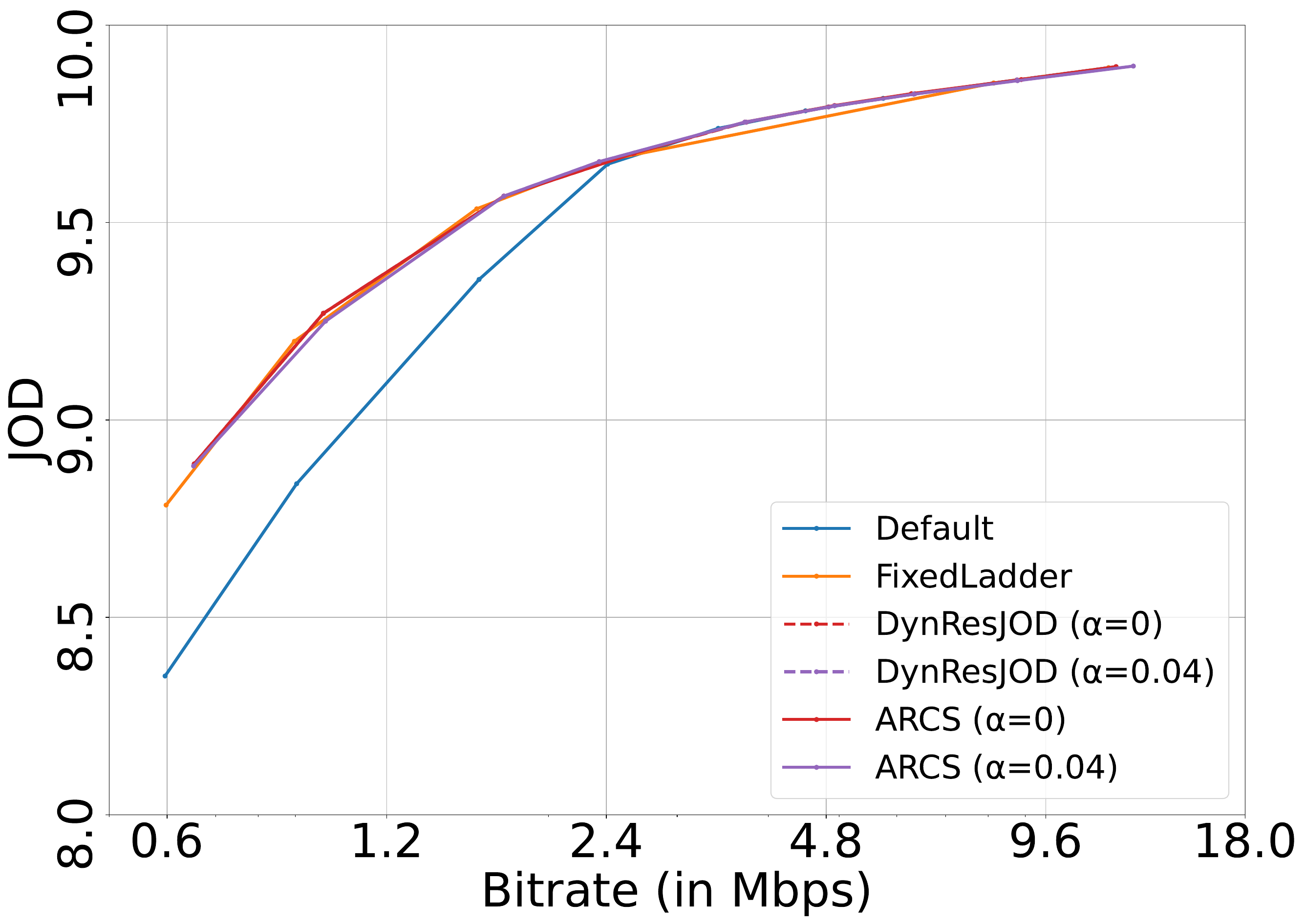}
\includegraphics[width=1\textwidth]{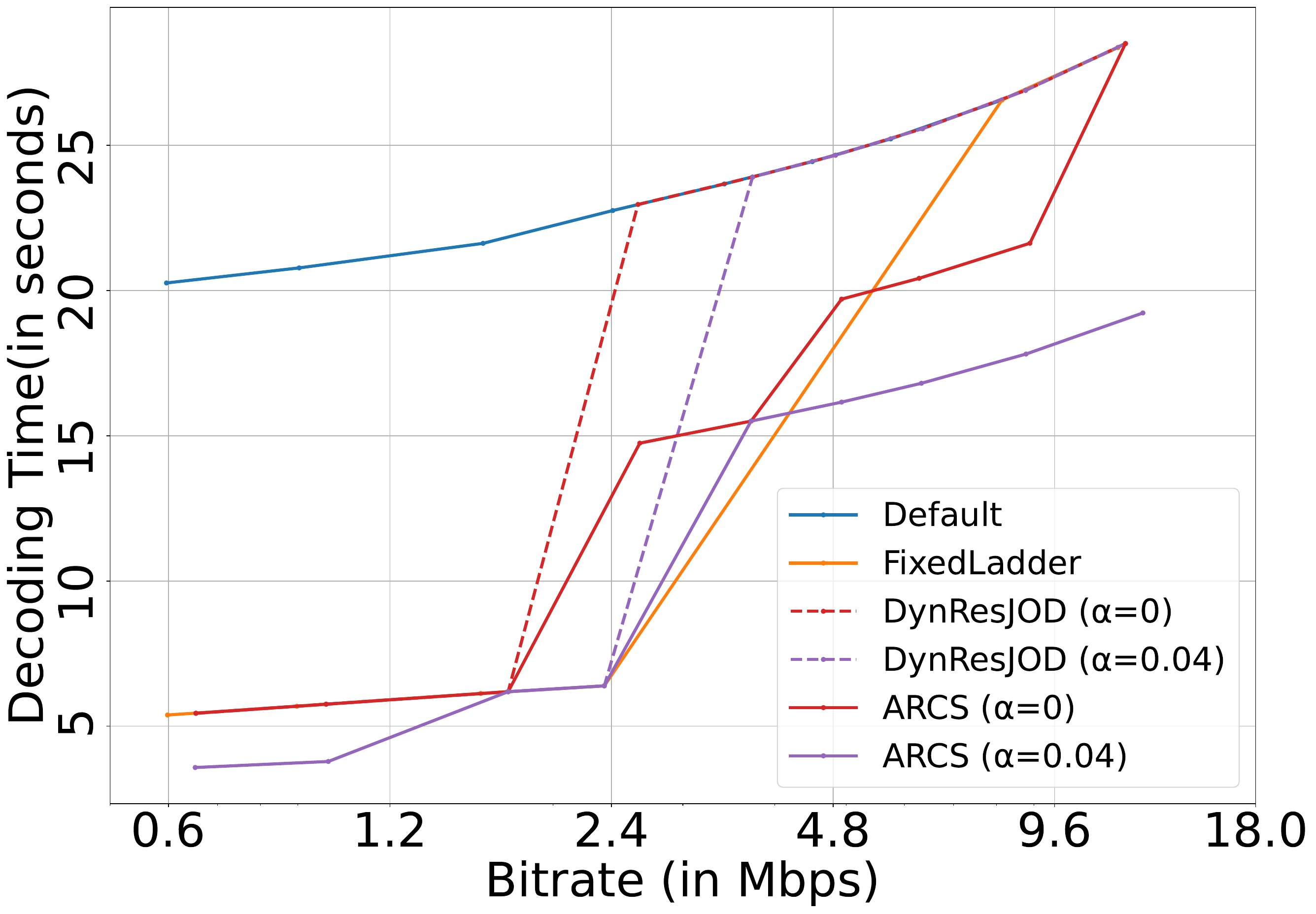}
\caption{Scarf}
\end{subfigure}
\begin{subfigure}{0.24\textwidth}
\centering
\includegraphics[width=1\textwidth]{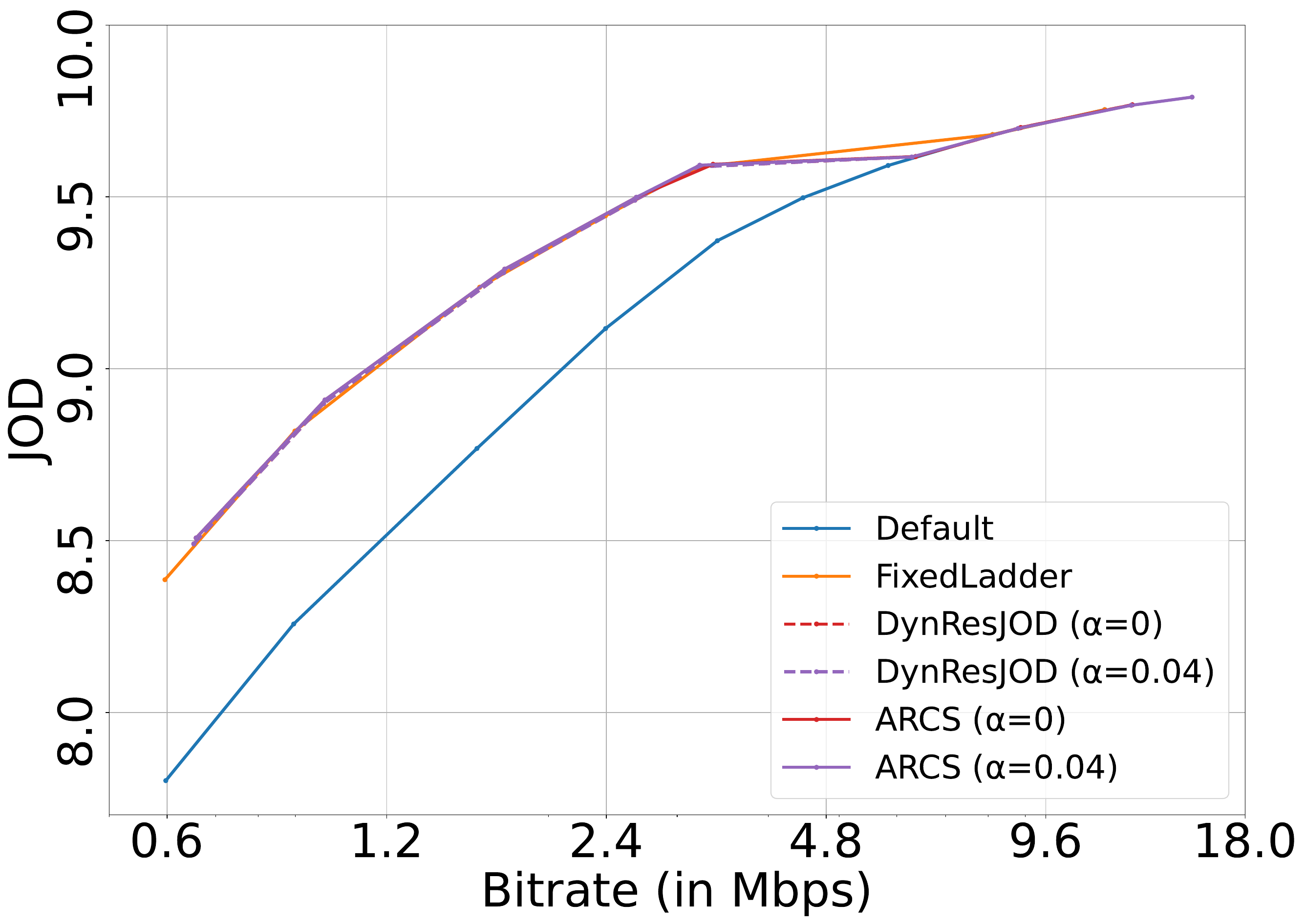}
\includegraphics[width=1\textwidth]{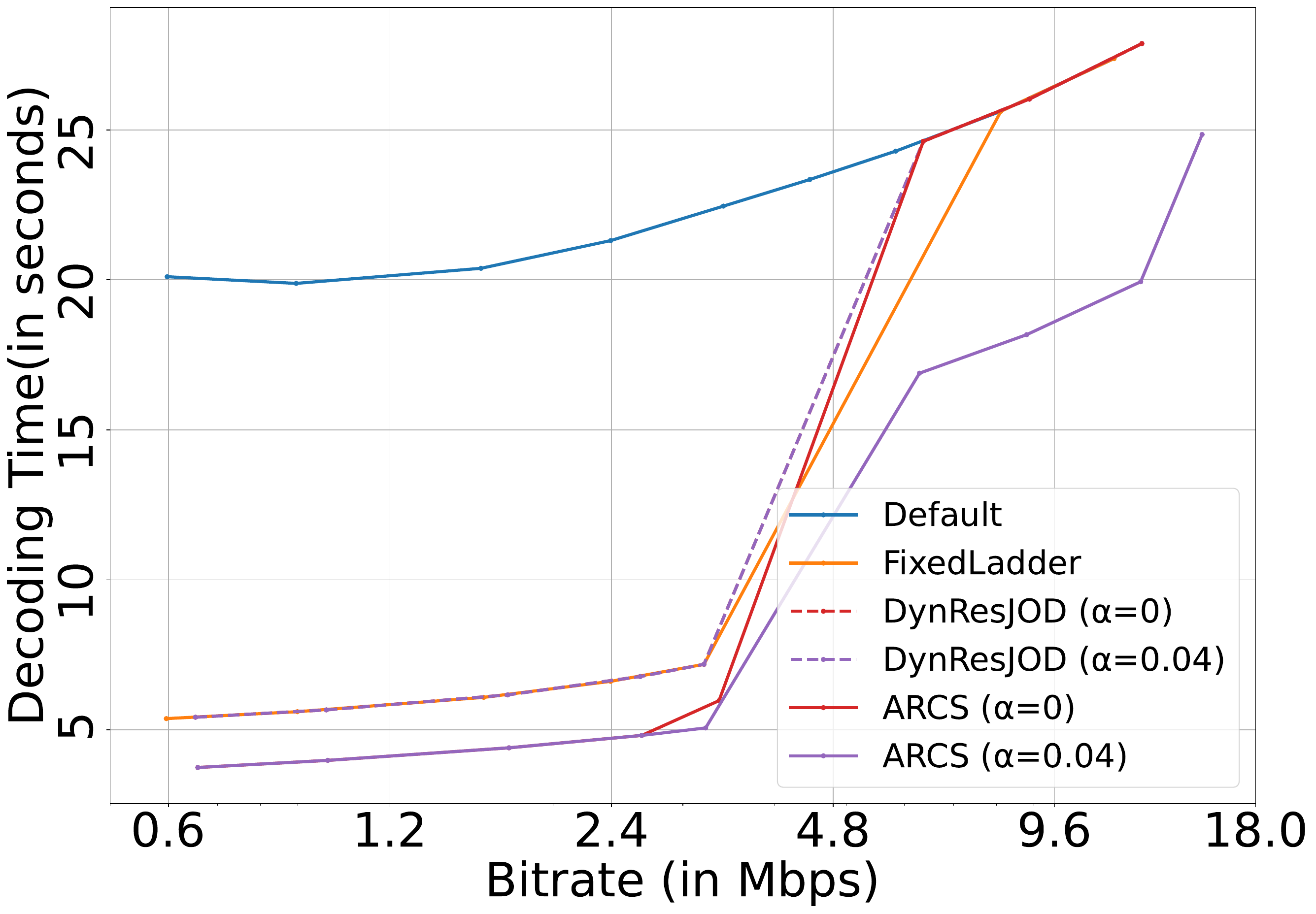}
\caption{Traffic Flow}
\end{subfigure}
\caption{Rate-colorVideoVDP, and rate-decoding time curves of two representative sequences.}
\label{fig:rd_res}
\end{figure}

Fig.~\ref{fig:rd_res} illustrates the rate–quality (top) and rate–decoding time (bottom) characteristics for the \emph{Scarf} and \emph{Traffic Flow} sequences under different configurations. The \emph{Default} setup, which encodes all representations at 2160p YUV444, achieves the highest ColorVideoVDP but incurs the largest decoding cost. In contrast, \emph{FixedLadder} exhibits reduced complexity but also apparent quality degradation at low bitrates due to fixed spatial scaling. \emph{DynResJOD} improves the trade-off by adaptively adjusting the spatial resolution, yet its lack of chroma adaptivity limits its efficiency, resulting in a plateau in quality improvement. The proposed \arcs\ framework achieves the most favorable rate–complexity balance. For both $\alpha=0$ and $\alpha=0.04$, it attains near-identical perceptual quality to the \emph{Default} configuration while reducing decoding time by up to \SI{50}{\percent}. The curves show a consistent downward shift in the rate–decoding time domain, confirming that joint resolution–chroma adaptation significantly lowers computational load without perceptual compromise.

\subsection{Effect of the Regularization Parameter $\alpha$}
\begin{figure}[t]
\centering
\begin{subfigure}{0.24\textwidth}
\centering
\includegraphics[width=\textwidth]{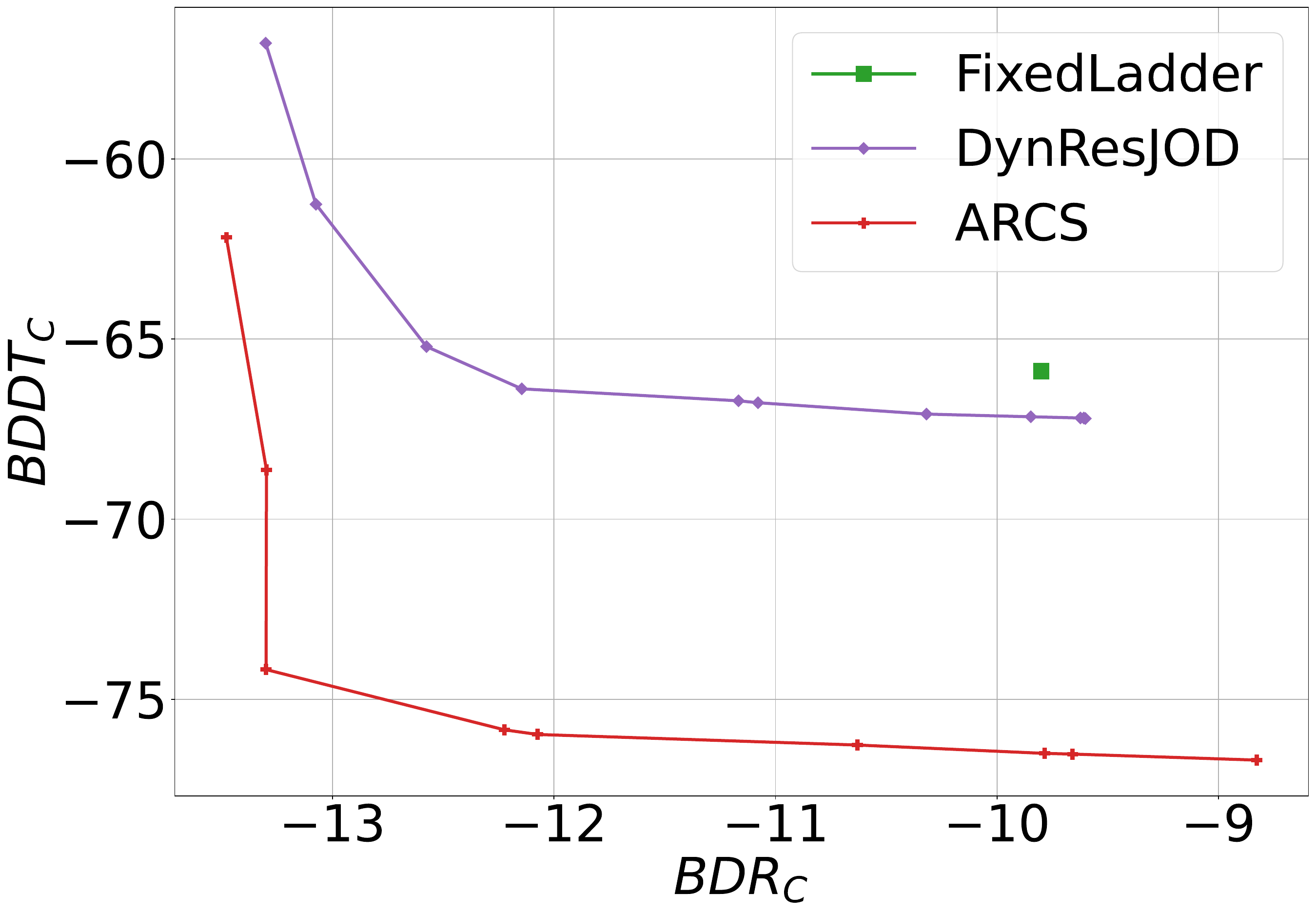}
\caption{BDR versus BDDT tradeoffs}
\end{subfigure}
\begin{subfigure}{0.24\textwidth}
\centering
\includegraphics[width=\textwidth]{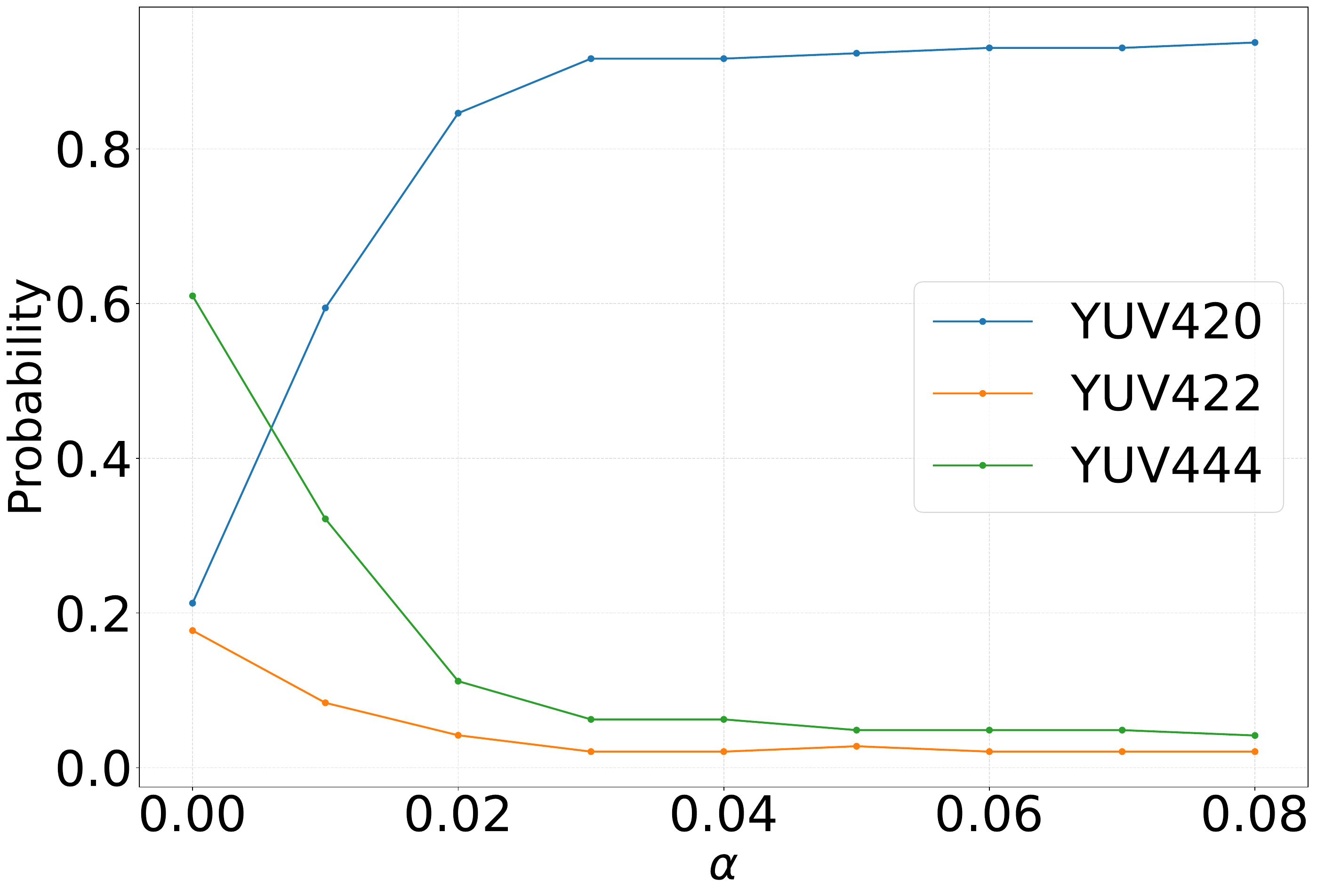}
\caption{PMF of colorspace}
\end{subfigure}
\caption{Results for various $\alpha$ values.}
\vspace{-1.4em}
\label{fig:alpha_sensibility}
\end{figure}

The regularization parameter $\alpha$ governs the balance between perceptual fidelity and decoding energy. As shown in Fig.~\ref{fig:alpha_sensibility}(a), increasing $\alpha$ progressively biases the optimization toward energy efficiency, moving along a smooth BDR–BDDT frontier. For small $\alpha$ values ($\alpha\leq0.02$), \arcs\ behaves as a conventional per-title optimizer, yielding up to \SI{13}{\percent} bitrate savings and around \SI{60}{}–\SI{70}{\percent} decoding-time reduction compared to \emph{Default}. When $\alpha$ increases to 0.04–0.08, decoding-time savings exceed \SI{75}{\percent} with only \SI{8}{}–\SI{12}{\percent} bitrate savings, marking the most balanced operation region. 

The probability mass function (PMF) in Fig.~\ref{fig:alpha_sensibility}(b) further illustrates this trend: as $\alpha$ rises, the share of YUV420 increases while YUV444 usage decreases, evidencing controlled chroma simplification under energy constraints. This gradual transition validates $\alpha$ as a stable and interpretable control factor for perceptual–energy trade-offs.


\subsection{Ablation Study: Impact of Chroma Adaptivity}
The contribution of chroma adaptivity is analyzed by comparing \emph{DynResJOD} with the proposed \arcs framework. Fig.~\ref{fig:alpha_sensibility}(a) presents the BDR–BDDT trade-offs for both methods across multiple $\alpha$ values. The \arcs configurations consistently occupy a more favorable region of the curve, indicating that for a given perceptual quality, \arcs achieves significantly lower decoding complexity. This demonstrates that integrating chroma adaptivity effectively shifts the efficiency frontier toward reduced energy consumption. The quantitative results in Table~\ref{tab:res} reinforce this observation. Across all $\alpha$ values, \arcs\ outperforms \emph{DynResJOD} in both bitrate efficiency and decoding-time reduction. At $\alpha=0.04$, for example, \arcs\ attains a \BDRC~of \SI{-12.07}{\percent}, and achieves up to \SI{75.97}{\percent} decoding-time savings ($BDDT_C$), outperforming \emph{DynResJOD} by about \SI{10}{\percent}. These results confirm that chroma adaptivity is the key enabler of the strong rate–energy trade-offs observed in \arcs.

\subsection{Overall Performance and Discussion}
Across benchmarks, \arcs\ delivers the most favorable rate–energy trade-off. Relative to \emph{Default}, \arcs\ achieves simultaneous coding and complexity gains: at $\alpha\!=\!0$, it reaches \SI{-13.48}{\percent} $\mathrm{BDR_C}$ with \SI{-62.18}{\percent} ($BDDT_C$) decoding-time reduction (Table~\ref{tab:res}). Increasing $\alpha$ steers the solution toward stronger efficiency: at $\alpha\!=\!0.04$, \arcs\ maintains near-baseline perceptual quality ($\mathrm{BDR_C}\!=\!\SI{-12.07}{\percent}$) while cutting decoding time by \SI{-75.97}{\percent}; at $\alpha\!=\!0.08$, decoding-time savings reach \SI{-76.68}{\percent}. The results show that introducing chroma as a first-class adaptation axis enables significant decoding-time reductions at negative BD-rate cost, extending conventional ABR design into the color domain for energy-efficient streaming.

\begin{table}[t]
\caption{Average performance of \arcs and benchmarks compared against \emph{Default}.}
\centering
\resizebox{0.855\linewidth}{!}{
\begin{tabular}{l|c|c|c|c|c}
\specialrule{.12em}{.05em}{.05em}
\specialrule{.12em}{.05em}{.05em}
\multicolumn{2}{c|}{\emph{Method}} & \BDRP & \BDRC & $BDDT_P$ & $BDDT_C$ \\
Name & $\alpha$ & [\%] & [\%] & [\%] & [\%]\\
\specialrule{.12em}{.05em}{.05em}
\specialrule{.12em}{.05em}{.05em}
\emph{FixedLadder} & - & 4.92 & -9.80 & -58.88 & -65.89\\
\hline
\multirow{5}{*}{DynResJOD} 
 & 0 & -7.21 & -13.30 & -49.05 & -56.79\\
 & 0.01 & -5.64 & -13.07 & -52.64 & -61.26\\
 & 0.02 & -4.81 & -12.58 & -56.60 & -65.21\\
 & 0.04 & -2.06 & -11.17 & -59.70 & -66.71\\
 & 0.08 & 1.81 & -9.62 & -61.04 & -67.19\\
\hline
\multirow{5}{*}{\arcs} 
 & 0 & -8.94 & -13.48 & -53.84 & -62.18\\
 & 0.01 & -1.45 & -13.30 & -62.39 & -68.63\\
 & 0.02 & 1.00 & -13.30 & -66.77 & -74.17\\
 & 0.04 & 4.15 & -12.07 & -69.21 & -75.97\\
 & 0.08 & 12.94 & -8.83 & -70.54 & -76.68\\
\specialrule{.12em}{.05em}{.05em}
\specialrule{.12em}{.05em}{.05em}
\end{tabular}
}
\vspace{-1.2em}
\label{tab:res}
\end{table}

\section{Conclusion}
We presented \emph{Adaptive Resolution–Chroma Subsampling} (\arcs), a framework that introduces chroma fidelity as a first-class adaptation axis alongside spatial resolution for energy-efficient video coding. By maximizing a composite quality–complexity objective under monotonicity constraints, \arcs\ constructs bitrate ladders that are both perceptually coherent and computationally efficient. On the SJTU UHD dataset, relative to \emph{Default}, \arcs\ attains simultaneous coding and complexity gains across $\alpha$: at $\alpha{=}0$, it achieves \SI{-13.48}{\percent} $\mathrm{BDR_C}$ with \SI{-62.18}{\percent} decoding-time reduction ($BDDT_C$); at $\alpha{=}0.04$, it maintains near-baseline perceptual quality ($\mathrm{BDR_C}{=}\SI{-12.07}{\percent}$) while cutting decoding time by \SI{-75.97}{\percent}; even at $\alpha{=}0.08$, decoding-time savings remain high (\SI{-76.68}{\percent}) with a favorable $\mathrm{BDR_C}$ of \SI{-8.83}{\percent}. The BDR–BDDT frontier show that increasing $\alpha$ smoothly shifts selections toward YUV420 when beneficial, confirming \arcs’ stable and interpretable control over the quality–energy trade-off. Practically, \arcs\ integrates naturally into per-title or per-segment workflows and is codec-agnostic (HEVC, VVC, AV1). Future work includes subjective validation, hardware-decoder energy profiling, end-to-end HAS integration with live ABR control, and learning-based prediction of resolution–chroma choices under device and content constraints.

\balance
\newpage
\bibliography{references.bib}{}
\bibliographystyle{IEEEtran}
\balance
\end{document}